\documentclass[%
aip,
jcp,
amsmath,amssymb,%
reprint,
]{revtex4-2}

\usepackage[pdftex]{graphicx}
\usepackage[version=3]{mhchem}
\usepackage{rotating}
\usepackage{multirow}
\usepackage{amsmath,newtxmath}
\usepackage{bm,braket}
\DeclareSymbolFont{cmlargesymbols}{OMX}{cmex}{m}{n}
\DeclareMathSymbol{\intop}{\mathop}{cmlargesymbols}{"5A}
  \def\int{\intop\nolimits}
\DeclareMathSymbol{\ointop}{\mathop}{cmlargesymbols}{"49}
  
\DeclareMathSymbol{\sumop}{\mathop}{cmlargesymbols}{"58}
  \let\sum\sumop
\DeclareMathSymbol{\prodop}{\mathop}{cmlargesymbols}{"59}
  \let\prod\prodop
\DeclareMathAlphabet{\mathcal}{OMS}{cmsy}{m}{n}

\newcommand{ \Eq   }[1]{Eq.~(\ref{#1})}
\newcommand{ \Eqs  }[2]{Eqs.~(\ref{#1}) and (\ref{#2})}

\newcommand{ \Table }[1]{Table \ref{tab:#1}}

\newcommand{ \Fig     }[1]{Fig.~\ref{fig:#1}}

\begin{document}
\title{Characterizing the embedded states of a fluorescent probe within a lipid bilayer using molecular dynamics simulations}
\author{Ryo Okabe}
\affiliation{Division of Chemical Engineering, Graduate School of Engineering Science, Osaka University, Toyonaka,
Osaka 560-8531, Japan}
\author{Natsuumi Ito}
\affiliation{Division of Chemical Engineering, Graduate School of Engineering Science, Osaka University, Toyonaka,
Osaka 560-8531, Japan}
\author{Yuya Matsubara}
\affiliation{Division of Chemical Engineering, Graduate School of Engineering Science, Osaka University, Toyonaka,
Osaka 560-8531, Japan}
\author{Nozomi Morishita Watanabe}
\affiliation{Division of Chemical Engineering, Graduate School of Engineering Science, Osaka University, Toyonaka,
Osaka 560-8531, Japan}
\author{Hiroshi Umakoshi}
\affiliation{Division of Chemical Engineering, Graduate School of Engineering Science, Osaka University, Toyonaka,
Osaka 560-8531, Japan}
\author{Kento Kasahara}
\email[Author to whom correspondence should be addressed: ]{kasahara@cheng.es.osaka-u.ac.jp}
\affiliation{Division of Chemical Engineering, Graduate School of Engineering Science, Osaka University, Toyonaka,
Osaka 560-8531, Japan}
\author{Nobuyuki Matubayasi}
\email{nobuyuki@cheng.es.osaka-u.ac.jp}
\affiliation{Division of Chemical Engineering, Graduate School of Engineering Science, Osaka University, Toyonaka,
Osaka 560-8531, Japan}

\begin{abstract}
The physicochemical properties of lipid bilayers (membranes) 
are closely associated with various cellular functions and 
are often evaluated using absorption and fluorescence spectroscopies.
For instance, by employing fluorescent probes that exhibit spectra reflective 
of the surrounding membrane environment, one can estimate the membrane polarity.
Thus, elucidating how such probes are embedded within the membranes would be 
beneficial for enabling a deeper interpretation of the spectra.
Here, we apply molecular dynamics (MD) simulation with an enhanced sampling method to 
investigate the embedded state of 6-propionyl-2-dimethylaminonaphthalene (Prodan) 
within a membrane composed of 1,2-dioleoyl-\textit{sn}-glycero-3-phosphocholine (DOPC), 
as well as its variation upon the addition of ethanol as a cosolvent to the aqueous phase.
In the absence of ethanol, it is found that the bulky moieties of Prodan (propionyl and dimethylamine groups) prefer to be oriented
toward the membrane center owing to the voids existing near the center.
The structural change in the membrane induced by the addition of ethanol
causes a reduction in the void population near the center,
resulting in a diminished orientation preference of Prodan.
\end{abstract}

\maketitle

\section{Introduction}
Lipids are the major components of cell membranes and form lipid bilayers.
The membrane properties are closely associated with cellular functions
and modulated by the lipid compositions.\cite{yeagle2004structure}
For instance, the membrane fluidity of a leukemic cell
is enhanced compared to that of the normal cell, and this difference
is attributed to the variation in lipid composition due to canceration.\cite{komizu2011membrane}
Elucidating the membrane properties is crucial also in terms of pharmacokinetics, because
drug permeability across the membranes correlates
with the membrane polarity and fluidity. \cite{liu2011lipophilicity, goldstein1984effects}
The membrane permeability varies depending on the cosolvents 
added to the aqueous phase that faces the membranes, 
through the structural change in the membrane.\cite{lopez2021effect} 
It is known that alcohols such as ethanol play a role as permeation enhancers 
by perturbing the membrane structures.\cite{pershing1990mechanism, rai2024alkanols}
To quantitatively evaluate the membrane properties mentioned above,
absorption and fluorescence spectroscopy\cite{lakowicz2006principles} are commonly employed.
In such measurements, fluorescent probes are embedded in the membranes,
exhibiting absorption and emission spectra that reflect the local membrane environment.
6-propionyl-2-dimethylaminonaphthalene (Prodan) 
and 6-dodecanoyl-2-dimethylaminonaphthalene (Laurdan)
are widely used probes for analyzing the interfacial and inner regions of the membranes, 
respectively.\cite{demchenko2009monitoring,gunther2021laurdan,krasnowska1998prodan, jurkiewicz2006headgroup} 
Owing to their high sensitivity to the surrounding environment,\cite{weber1979synthesis, macgregor1986estimation, massey1998surface, krasnowska2001surface}
these probes exhibit characteristic Stokes shifts that reflect local polarity 
and hydrogen-bonding interactions within the membrane.
The local membrane polarity is often quantified using the fluorescence intensities 
at specific wavelengths in the spectrum, expressed as the generalized polarity (GP).
\cite{parasassi1991quantitation, parasassi1995abrupt, parasassi1995membrane, parasassi1998laurdan} 
Very recently,
a scheme in systematically analyzing distinct regions within the membrane
was constructed based on the fluorescence decays in a series of solvent mixtures.
\cite{ito2023multiplicity, ito2024multifocal}
In the spectroscopy-based approach,
the embedded states of the probes are represented by their depth within the membrane.
Thus, elucidating the embedded states at the atomistic level 
would be useful for gaining further physicochemical insights from the spectra.

Molecular dynamics (MD) simulation is a representative computational method 
to analyze systems of interest in atomistic detail.\cite{allen1989computer, frenkel2001understanding}
Advances in computational power have made it possible
to explore complex systems, including membrane systems.\cite{jurkiewicz2012lipid, Moradi_2019, hsieh2021all, majumder2024probing}
Prodan and Laurdan have been extensively investigated using MD simulations 
as the fluorescent probes embedded in the membranes.
\cite{cwiklik2011absorption, nitschke2012molecular,suhaj2018prodan,suhaj2020laurdan}
The quantum mechanics/molecular mechanics (QM/MM) method,
combined with molecular dynamics (MD) simulations, is also utilized
to analyze the relationship between the embedded states of probes (structure, position, and orientation) 
and its spectra.\cite{osella2016investigation,wasif2018orientation, osella2021influence,knippenberg2024hydration}
In the above studies, the analysis was done for the specific embedded states.  
Given that the experimentally observed spectra reflect different embedded states weighted with the Boltzmann factor,
elucidating the probe's embedded states in terms of thermodynamics 
would enable a deeper interpretation of the spectra. 

In the present study,
we elucidate the embedded state of Prodan based on the free-energy landscape 
by using the replica-exchange umbrella sampling (REUS) MD simulation.
\cite{sugita1999replica, sugita2000multidimensional, murata2004free}
The REUS method is an enhanced sampling technique 
that enables an efficient sampling of the configurations 
along a predefined coordinate called the reaction coordinate, 
by employing multiple replicas of the system of interest 
under different biased potentials. 
In the membrane systems, 
enhanced sampling techniques including the REUS method 
have been extensively applied to study drug permeation through the membranes 
for computing the free-energy landscape along a permeation pathway.
\cite{Venable_2019,bemporad2004permeation,ghaemi2012novel,lee2016simulation, cipcigan2020membrane, sugita2021large}
Thus, such methods are expected to provide reliable estimations of the embedded states of probes. 

We apply the REUS method to membrane–probe systems in which Prodan is embedded 
in membranes composed of 1,2-dioleoyl-\textit{sn}-glycero-3-phosphocholine (DOPC), 
facing different aqueous phases. 
As aqueous phases, we examine pure water, 1 M ethanol, and 2 M ethanol solutions.
It is well known that the addition of alcohol as a cosolvent induces the structure change in the membranes 
\cite{rai2024alkanols,matsubara2024methodology,gurtovenko2009interaction, barry1995effects, rottenberg1992probing}, 
leading to the change in the spectra.\cite{zeng1995effect}
Hence, we discuss how the embedded states are altered upon ethanol addition.
Subsequently, the distribution of voids within the membrane and its variation upon ethanol addition are analyzed, as these voids may stabilize specific embedded states through excluded-volume effects.
Since the electrostatic interaction energy between Prodan and its surrounding environment predominantly contributes to its absorption and emission spectra, the interaction energies are also evaluated across different states.
\section{Methods}
In this section, we briefly describe the replica-exchange umbrella sampling (REUS) method. \cite{sugita2000multidimensional}
Let us consider a membrane–probe system whose phase-space coordinate is denoted by $\bm{\Gamma}$,
with a potential energy function $U\left(\bm{\Gamma}\right)$.
In the REUS method, multiple replicas of the system are simulated,
each subject to a different biased potential applied along a predefined reaction coordinate $\zeta$.
The total number of replicas is denoted by $K$, and the biased potential
applied to the $i$th replica is given by $V_i\left(\zeta\right)$.
The functional form of $V_i\left(\zeta\right)$ is defined as
\begin{align}
V_{i}\left(\zeta\right) & =k_{i}\left(\zeta-\zeta_{i}\right)^{2}, \label{Vi}
\end{align}
where $k_i$ and $\zeta_i$ represent the force constant and the center of the bias potential, respectively.
The modified potential energy of the $i$th replica, $U_{i}\left(\bm{\Gamma}\right)$, is expressed as
\begin{align}
U_{i}\left(\bm{\Gamma}\right) & =U\left(\bm{\Gamma}\right)+V_{i}\left(\zeta\right).
\end{align}
During the REUS simulation,
the configurations are stochastically exchanged between replicas with adjacent indices 
according to the Metropolis algorithm.
This exchange allows the system to escape local minima, thereby 
accelerating convergence in the free-energy calculation. 

The ensemble averages in the unbiased system can be evaluated 
using the multistate Bennett acceptance ratio (MBAR) method\cite{shirts2008statistically}, 
based on the configurations generated from the REUS simulation, as described below.
Let ${\bf \Gamma}_{in}~ (n=1,2,\cdots, N_{i})$ represent the $n$th configuration 
in the $i$th replica and $N_{i}$ be the number of configurations in the $i$th replica.
The free energy of the $i$th replica, $f_{i}$, is expressed with
\begin{align}
f_{i} & =-\dfrac{1}{\beta}\log\left(\sum_{j=1}^{K}\sum_{n=1}^{N_{i}}\dfrac{\exp\left[-\beta U_{i}\left(\bm{\Gamma}_{jn}\right)\right]}{{\displaystyle \sum_{k=1}^{K}N_{k}\exp\left[\beta\left(f_{k}-U_{k}\left(\bm{\Gamma}_{jn}\right)\right)\right]}}\right), 
\end{align}
where $\beta$ is the inverse temperature.
Then, the statistical weight of each configuration, $W\left(\bm{\Gamma}_{in}\right)$, can be 
expressed using $f_{i}$ as 
\begin{align}
W\left(\bm{\Gamma}_{in}\right) & =\dfrac{1}{C}\dfrac{\exp\left[\beta\left(U\left(\bm{\Gamma}_{in}\right)\right)\right]}{{\displaystyle \sum_{k=1}^{K}}N_{k}\exp\left[\beta\left(f_{k}-U_{k}\left(\bm{\Gamma}_{in}\right)\right)\right]}.
\end{align}
Here, $C$ is the normalization constant that ensures 
\begin{align}
\sum_{i=1}^{K}\sum_{n=1}^{N_{i}}W\left(\bm{\Gamma}_{in}\right) & =1.
\end{align}
Using $W\left(\bm{\Gamma}_{in}\right)$, one can estimate the ensemble 
average of arbitrary quantity $A$, $\braket{A}$, as
\begin{align}
 \braket{A} = \sum_{i=1}^{K}\sum_{n=1}^{N_{i}} A\left(\bm{\Gamma}_{in}\right)W\left(\bm{\Gamma}_{in}\right). 
\end{align} 
\section{Computational details}
\subsection{Simulation setups}
\begin{figure}[t]
  \centering
  \includegraphics[width=0.9\linewidth]{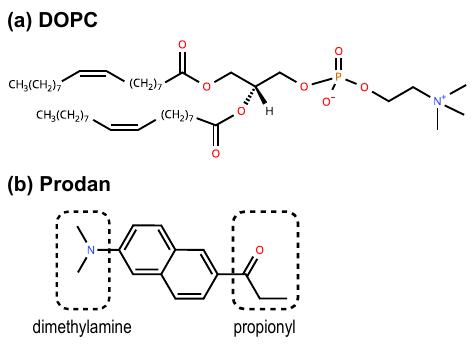}
  \caption{Chemical structures of (a) 1,2-dioleoyl-\textit{sn}-glycero-3-phosphocholine (DOPC) and (b) 6-propionyl-2-dimethylaminonaphthalene (Prodan).}
  \label{fig:fig1}
\end{figure}

We investigated the membrane-probe systems composed of 
1,2-dioleoyl-\textit{sn}-glycero-3-phosphocholine (DOPC) lipid bilayer and a  
6-propionyl-2-dimethylaminonaphthalene (Prodan) (\Fig{fig1}).
For the aqueous solutions facing the membrane, 
pure water, 1 M ethanol, and 2 M ethanol aqueous solutions 
were used. 
The force fields for DOPC, Prodan, ethanol, and water were CHARMM36,\cite{klauda2010update} 
CHARMM generalized force field (CGenFF),\cite{vanommeslaeghe2010charmm} 
CGenFF, and CHARMM-compatible TIP3P,\cite{jorgensen1983comparison} 
respectively.
In this study, a pH of 7.0 is assumed, under which the dimethylamine group in Prodan 
is considered to be unprotonated (neutral), given that the $\mathrm{p}K_{\mathrm{a}}$ of the dimethylamine group  
is $\sim${}$4.5$.\cite{chong1989effects}
The electronic structure of Prodan was calculated 
using the CAM-B3LYP/cc-pVDZ level calculation.\cite{yanai2004new,dunning1989gaussian}
Then, the charges from electrostatic
potentials using a grid based method (CHelpG) was used 
for calculating the point charges on the atoms in Prodan.\cite{breneman1990determining}
The above quantum chemical calculation was performed 
using Gaussian 16.\cite{frisch2016gaussian} 
The target temperature and pressure were set to 310 K and 1 atm, respectively.

The initial configurations of the membrane were prepared using CHARMM-GUI server.
\cite{jo2008charmm,wu2014, lee2018charmm}
The number of DOPC molecules were 128 (64 per leaflet) for all the systems.
The numbers of water and ethanol molecules, $\left(N_{\mathrm{wat}}, N_{\mathrm{etoh}}\right)$, 
were $\left(7680, 0\right)$ for pure water, $\left(7250, 140\right)$ for 1 M ethanol, 
and $\left(6910, 280\right)$ for 2 M ethanol.
The initial configurations of the aqueous phases were built using Packmol.\cite{martinez2009packmol} 

All the MD simulations were performed with GENESIS 2.0.\cite{jung2015genesis,kobayashi2017genesis,jung2021new}
We used the Bussi method for controlling the temperature and pressure.\cite{Bussi_2007}
For the initial equilibration stage, we performed MD simulations (1.875 ns in total; $NVT$ and $NPT$) 
using the velocity Verlet (VVER) integrator\cite{swope1982computer} with a time step of 2 fs, 
applying restraints to the membranes in accordance with the CHARMM-GUI guidelines.
Subsequently, the reversible reference system propagator algorithm (r-RESPA) with a time interval of 2.5 fs
was used for the simulations under $NPT$ condition.
We conducted the 550 ns MD simulation for further equilibration.
After the equilibration, we performed the REUS simulations with 32 replicas.
In the REUS simulations, the reaction coordinate was defined as the $z$-component of the center of mass (CoM) of Prodan, 
where the $z$-direction is normal to the membrane surface and $z = 0$ corresponds to the CoM of the membrane.
Performing 30 ns MD simulations with automatic tuning of the REUS parameters,\cite{bonomi2019biomolecular}
the reference positions ${z_i}$ (corresponding to $\zeta_i$ in \Eq{Vi}) and the force constants ${k_i}$
were optimized (Tables S1--S4 of the supplementary material).
After the 30 ns simulations without exchange, we conducted 500 ns REUS simulations for production.
We also performed two independent 500 ns 
REUS simulations using the tuned parameters described above
for estimating the statistical error. 
In these runs, 550 ns MD simulation was conducted for equilibration prior to the REUS simulations, 
starting from different initial configurations.
The weights of the configurations in the REUS trajectories were computed 
by means of the MBAR method implemented in GENESIS.\cite{matsunaga2022use}
The interaction energy analysis were performed using ERmod 0.3.7.\cite{Sakuraba_2014} 

To analyze the void existing inside the membrane,
we performed the MD simulations for the membrane systems in the absence of Prodan.
The initial configurations were prepared from those of the membrane-probe systems 
obtained after 550 ns equilibration mentioned above.
After deleting the Prodan's coordinates, the energy minimization was conducted. 
Then, we performed 100 ps NVT simulations using VVER, followed by 100 ns NPT simulations using r-RESPA.
The last 50 ns were used for analysis. 
\section{Results and discussion}
\subsection{Free-energy landscapes}
\begin{figure*}[t]
  \centering
  \includegraphics[width=1.0\linewidth]{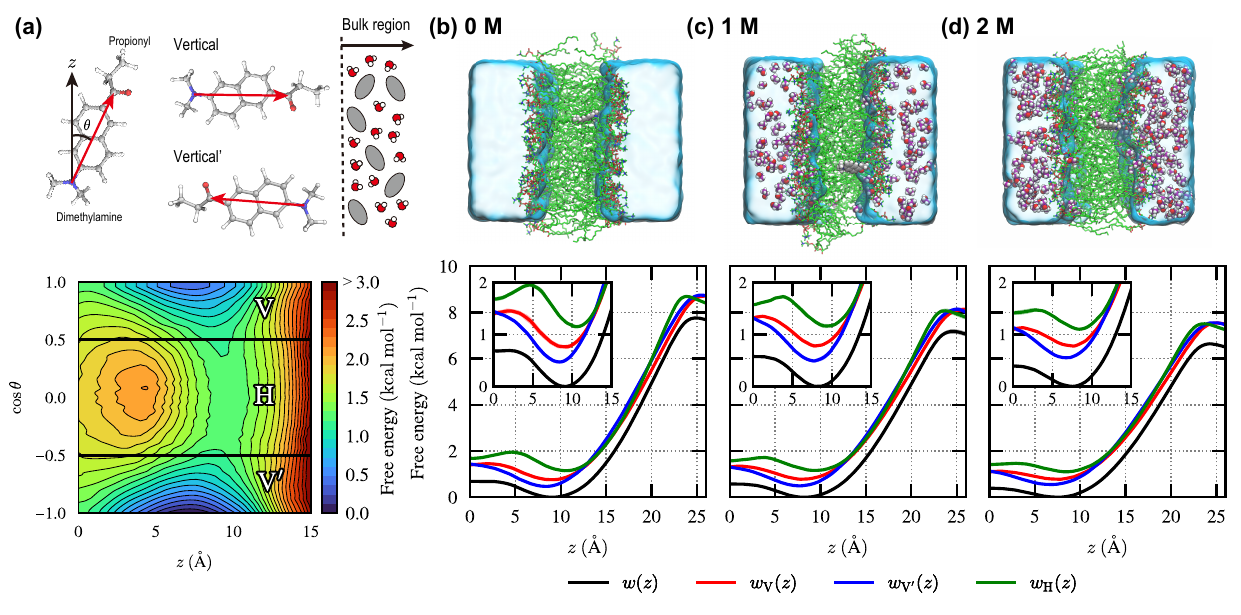}
  \caption{Free-energy landscapes (FELs) for the embedded states of Prodan. (a) schematic illustration of Prodan's orientation (top), $\theta$,  and the 2D-FEL in pure water along the $z$-component of CoM for Prodan, $z$, and  $\cos{\theta}$ (bottom), (b)$-$(d) 1D-FELs along $z$, $w\left(z\right)$ (\Eq{wz}), and their decomposition into different regions, $w_{\mathrm{X}}\left(z\right)$ ($\mathrm{X=V, V^{\prime},~or~ H}$) (\Eq{wXz}), for the 0 (pure water), 1 M, and 2 M ethanol cases, respectively.}
  \label{fig:fig2}
\end{figure*}
We first examine the two-dimensional free-energy landscape (2D-FEL)
as a function of the $z$-component of the center of mass (CoM) of Prodan, $z$,
and the cosine of the tilt angle, $\cos\theta$ (\Fig{fig2}(a)).
The tilt angle, $\theta$, is defined as the angle between the membrane normal (the $z$-direction)
and the Prodan's molecular axis pointing from the nitrogen atom in the dimethylamine group
to the carbonyl carbon atom in the propionyl group.
It can be seen from the 2D-FEL for the 0 M ethanol (pure water) case
that stable embedded states appear around $z \sim 8~\mathrm{\AA}$,
with the molecular axis nearly parallel to the $z$-direction ($\cos \theta \sim \pm 1$).
On the other hand, the inner region ($z < 4~\mathrm{\AA}$)
is found to be unfavorable for Prodan especially
when its molecular axis is oriented orthogonal to the $z$-direction.
Then, we define three regions based on $\cos \theta$ as follows:
$\cos \theta > 0.5$, $\cos\theta < -0.5$,  and $-0.5 \leq \cos \theta \leq 0.5$
are referred to as vertical ($\mathrm{V}$), vertical' ($\mathrm{V}^{\prime}$), and horizontal ($\mathrm{H}$), respectively.
Since similar profiles are observed in the 2D-FELs for the 1 M and 2 M ethanol cases 
(Fig. S1 of the supplementary material),
this definition is useful for analyzing how the thermodynamic stability of each region changes in the presence of ethanol. 

To closely examine the change in thermodynamic stability for each region,  
we introduce the one-dimensional FELs (1D-FELs)  
with and without restriction to a specific region $\mathrm{X}$  
($\mathrm{X} = \mathrm{V}, \mathrm{V}^{\prime}$, or $\mathrm{H}$), defined respectively as

\begin{align}
  w_{\mathrm{X}}\left(z\right) & =-\dfrac{1}{\beta}\log\dfrac{\Braket{\delta\left(z-z_{\mathrm{s}}\left(\bm{\Gamma}\right)\right)\Theta_{\mathrm{X}}\left(\cos\theta\right)}}{\Braket{\Theta_{\mathrm{X}}\left(\cos\theta\right)}}+C, \label{wXz} \\
  w(z) 
  &= -\dfrac{1}{\beta} \log \Braket{ \delta\left( z - z_{\mathrm{s}}(\bm{\Gamma}) \right) } + C, \label{wz}
\end{align}
where $z_{\mathrm{s}}(\bm{\Gamma})$ is the $z$-component of CoM of Prodan (solute) 
for a given phase-space coordinate $\bm{\Gamma}$,  
and $\Theta_{\mathrm{X}}(\cos \theta)$ is the characteristic function  
that equals unity when $\cos \theta \in \mathrm{X}$ and zero otherwise.
$C$ is the normalization constant, which is determined such that 
the lowest value of $w\left(z\right)$ is zero.
\Fig{fig2}(b)--(d) illustrate the 1D-FELs for the 0, 1, and 2 M ethanol cases, respectively.
For all the cases,
region $\mathrm{V}^{\prime}$ exhibits 
the most stable state at $\sim${}$8~\mathrm{\AA}$.
It is interesting to note that region $\mathrm{V}^{\prime}$ is more stable than region $\mathrm{V}$, despite the stabilization effect from the electrostatic interaction between Prodan and the aqueous phase being weaker in region $\mathrm{V}^{\prime}$ than in region $\mathrm{V}$ due to the increased distance between the propionyl group and the aqueous phase in region $\mathrm{V}^{\prime}$ (\Fig{fig2}(a)).
Region $\mathrm{H}$ at the corresponding $z$ is found to be unstable compared with regions $\mathrm{V}$ and $\mathrm{V}^{\prime}$, as is also seen in the 2D-FEL (\Fig{fig2}(a)).
As the ethanol concentration increases, 
the difference in thermodynamic stability among the different regions at $z \leq 15~\mathrm{\AA}$ becomes smaller, 
indicating that the extent of the orientation preference is mitigated by adding ethanol.
Regarding the interface ($z\sim 20-25~\mathrm{\AA}$), 
a high free-energy barrier is present, 
reflecting the hydrophobic nature of Prodan, 
and the difference in the barrier height between different regions is negligibly small.
In the presence of ethanol, it is observed that 
the barrier height is lowered.
This trend is consistent with experimental observations 
that ethanol acts as a permeation enhancer.\cite{pershing1990mechanism, rai2024alkanols} 
\begin{figure*}[t]
  \centering
  \includegraphics[width=0.8\linewidth]{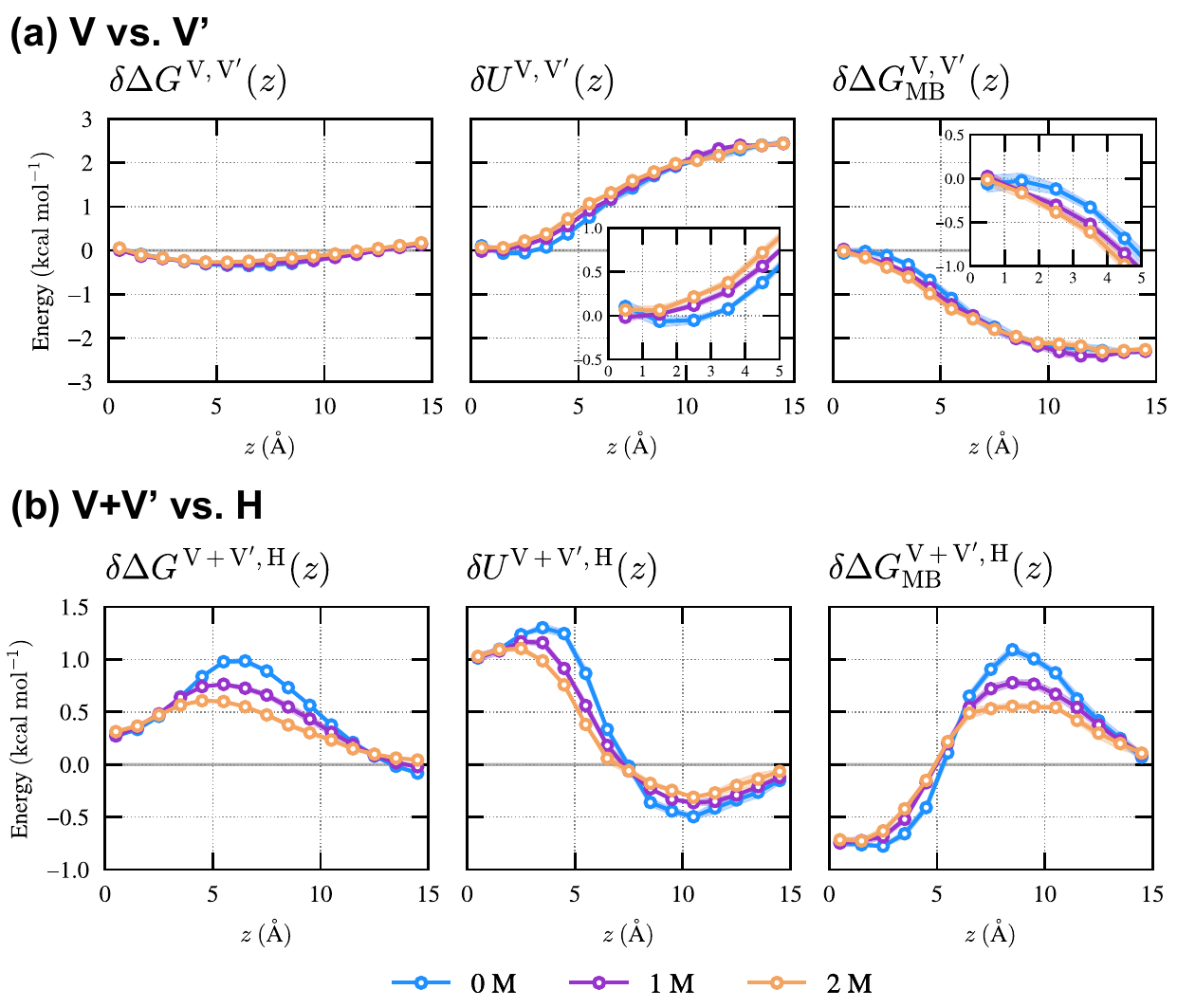}
  \caption{Differences in the 1D-FELs between different regions, $\delta \Delta G^{\mathrm{X}_{0},\mathrm{X}_{1}}$, and their decomposition into the difference in the interaction energy of Prodan with surrounding environments, $\delta U^{\mathrm{X_0,X_1}}\left(z\right)$, and differences in the many-body entropic contribution, $\delta \Delta G_{\mathrm{MB}}^{\mathrm{X_0,X_1}}\left(z\right)$. 
(a) $\mathrm{X}_{0}=\mathrm{V},\mathrm{X}_{1} = \mathrm{V}^{\prime}$, (b) $\mathrm{X}_{0}=\mathrm{V+V^{\prime}}, \mathrm{X}_{1} = \mathrm{H}$. The left, middle, and right panels of (a) and (b) show the profiles of $\delta \Delta G^{\mathrm{X_{0},X_{1}}}\left(z\right)$, $\delta U^{\mathrm{X_{0},X_{1}}}\left(z\right)$, and $\delta \Delta G_{\mathrm{MB}}^{\mathrm{X_{0},X_{1}}}\left(z\right)$, respectively.
\label{fig:fig3}} 
\end{figure*}
\subsection{Decomposition of free energy}
In this subsection, we discuss the change 
in the stability of each state due to the addition of ethanol in terms of
the systematic decomposition of free energy according to the classical density functional theory.  
The free-energy difference between regions 
$\mathrm{X}_{0}$ and $\mathrm{X}_{1}$, $\delta \Delta G^{\mathrm{X_{0},X_{1}}}\left(z\right)$, 
given by
\begin{align}
\delta\Delta G^{\mathrm{X_{0},X_{1}}}\left(z\right) & =-\dfrac{1}{\beta}\log\dfrac{\Braket{\Theta_{\mathrm{X_{1}}}\left(\cos\theta\right)\delta\left(z-z_{\mathrm{s}}\left(\bm{\Gamma}\right)\right)}}{\Braket{\Theta_{\mathrm{X_{0}}}\left(\cos\theta\right)\delta\left(z-z_{\mathrm{s}}\left(\bm{\Gamma}\right)\right)}},
\end{align}
can be decomposed into the difference in the interaction energy of Prodan 
with the surrounding environment, $\delta U^{\mathrm{X_{0},X_{1}}}\left(z\right)$, 
and the many-body entropic contribution, $\delta \Delta G_{\mathrm{MB}}^{\mathrm{X}_{0},\mathrm{X}_{1}}\left(z\right)$, as
\begin{align}
\delta\Delta G^{\mathrm{X_{0},X_{1}}}\left(z\right) & =\delta U^{\mathrm{X_{0},X_{1}}}\left(z\right)+\delta\Delta G_{\mathrm{MB}}^{\mathrm{X_{0},X_{1}}}\left(z\right).
\end{align} 
$\delta \Delta G_{\mathrm{MB}}^{\mathrm{X}_{0},\mathrm{X}_{1}}\left(z\right)$ is the difference 
in the free-energy penalty 
due to the distortion of the surrounding environment upon 
the insertion of Prodan between regions $\mathrm{X}_{0}$ and $\mathrm{X}_{1}$.

\Fig{fig3}(a)
shows the free-energy difference between regions $\mathrm{V}$ and $\mathrm{V}^{\prime}$, 
denoted as $\delta \Delta G^{\mathrm{V},\mathrm{V}^{\prime}}\left(z\right)$, together with its decomposition.
It is observed that the slightly higher stability of region $\mathrm{V}^{\prime}$ 
compared to region $\mathrm{V}$  
stems from a more negative value of $\delta \Delta G_{\mathrm{MB}}^{\mathrm{V},\mathrm{V}^{\prime}}$, suggesting that the extent of membrane distortion upon Prodan insertion  
is smaller in region $\mathrm{V}^{\prime}$ than in region $\mathrm{V}$, especially when Prodan is distant from the membrane center. 
The positive value of $\delta U^{\mathrm{V,V^{\prime}}}\left(z\right)$ 
indicates that the stabilization effect of the interaction 
between Prodan and its surrounding environment is stronger in region $\mathrm{V}$ than 
in region $\mathrm{V}^{\prime}$, 
although the magnitude of $\delta U^{\mathrm{V,V^{\prime}}}\left(z\right)$ 
is smaller than that of $\delta \Delta G_{\mathrm{MB}}^{\mathrm{V,V^{\prime}}}\left(z\right)$. 
As the ethanol concentration increases, the stability of region $\mathrm{V}^{\prime}$  
relative to region $\mathrm{V}$ slightly decreases due to an increase in $\delta U^{\mathrm{V},\mathrm{V}^{\prime}}\left(z\right)$.  

The stability of region $\mathrm{H}$ is lower than
that of the composite region of
$\mathrm{V}$ and $\mathrm{V}^{\prime}$, $\mathrm{V}+\mathrm{V}^{\prime}$, for the 0 M ethanol case,
as indicated by $\delta \Delta G^{\mathrm{V}+\mathrm{V}^{\prime},\mathrm{H}}\left(z\right)$ 
(\Fig{fig3}(b)), especially at $z\sim${}$6~\mathrm{\AA}$.
Interestingly, the decomposition of $\Delta G^{\mathrm{V+V^{\prime},H}}\left(z\right)$ reveals that 
the factors contributing to the stabilization or destabilization of region H with respect to region $\mathrm{V+V^{\prime}}$ vary depending on the $z$-range.
The destabilizing effect of $\delta U^{\mathrm{V+V^{\prime},H}}\left(z\right)$ 
is evident for $z\lesssim 7.5~\mathrm{\AA}$ and reaches its maximum 
around $z\sim${}$4~\mathrm{\AA}$, resulting in the most unstable state in region $\mathrm{H}$ at $z\sim${}$6~\mathrm{\AA}$.
It is seen that $\delta \Delta G_{\mathrm{MB}}^{\mathrm{V+V^{\prime},H}}\left(z\right)$ contributes to the stabilization for $z\lesssim 5~\mathrm{\AA}$ and its magnitude is smaller than that of $\delta U^{\mathrm{V+V^{\prime},H}}\left(z\right)$.
$\Delta G_{\mathrm{MB}}^{\mathrm{V+V^{\prime},H}}\left(z\right)$, in turn, significantly 
destabilizes region $\mathrm{H}$ around $z\sim${}$8~\mathrm{\AA}$.
In the presence of ethanol, region H is found to be stabilized due to the reduction in the destabilizing effects of $\delta U^{\mathrm{V+V^{\prime},H}}\left(z\right)$ and $\delta \Delta G_{\mathrm{MB}}^{\mathrm{V+V^{\prime},H}}\left(z\right)$. 
\subsection{Void-size distribution along $z$-direction\label{sec:vsd}}
\begin{figure*}[t]
  \centering
  \includegraphics[width=1.0\linewidth]{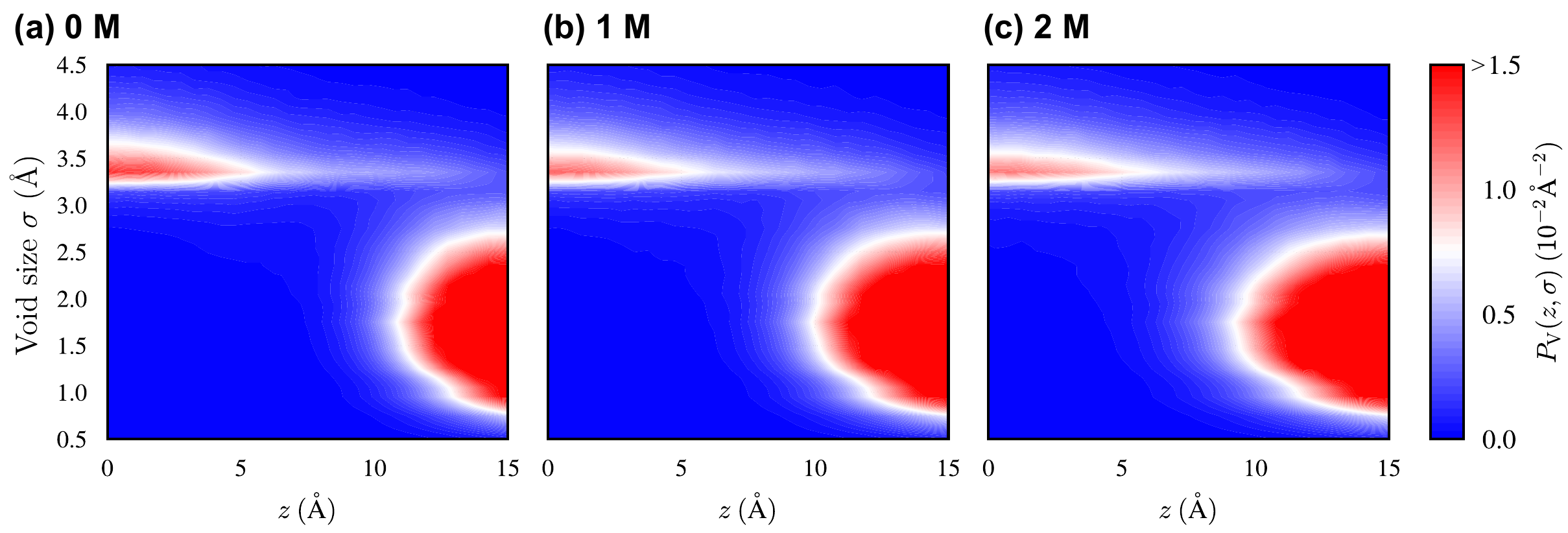}
  \caption{Void-size distributions along the $z$-direction, $P\left(z,\sigma\right)$, for (a) 0, (b) 1, and (c) 2 M ethanol cases.\label{fig:fig4}}
  
\end{figure*}
We discuss the structural property of the membrane.
According to previous simulation studies on membrane permeation, 
\cite{shinoda2004molecular,cardenas2014modeling,chipot2016subdiffusion} 
voids, defined as spaces not occupied by atoms,  
exist within the membrane, particularly near its center.
Such voids can reduce the free-energy penalty  
due to the distortion of the membrane caused 
by the embedding of molecules that are related with $\delta \Delta G^{\mathrm{X}_{0},\mathrm{X}_{1}}_{\mathrm{MB}}$.
Therefore, quantifying these voids is essential 
for elucidating the embedded states.
In this work, we compute
the distribution of the void size ($\sigma$) along the $z$-direction, $P_{\mathrm{V}}\left(z,\sigma\right)$,  
based on an algorithm developed in the field of computational material science (Appendix \ref{sec:def_vsd}).\cite{sarkisov2020materials}
The trajectory in the absence of Prodan is used for computing $P_{\mathrm{V}}\left(z,\sigma\right)$. 

\Fig{fig4} shows the $P_{\mathrm{V}}\left(z,\sigma\right)$ for the 0, 1, and 2 M ethanol cases.
In all cases, large voids ($\sigma \geq 3~\mathrm{\AA}$) are populated near the membrane center.
This observation is consistent with the previous studies.\cite{cardenas2014modeling,chipot2016subdiffusion}
Such large voids can accommodate bulky molecules or moieties with a low free-energy penalty.
Thus, the stabilizing effect of the voids on region $\mathrm{H}$
is pronounced for the many-body entropic term when Prodan is located near the membrane center,
as all of its atoms are embedded within the membrane center.
A broad distribution of $\sigma$ is observed around $z=12-15\mathrm{\AA}$, where the glycerol groups of DOPC 
molecules are located (Fig. S2 of the supplementary material).
Since the length of Prodan along its molecular axis is $\sim${}$10~\mathrm{\AA}$,  
the presence of voids near the membrane center and $z=12-15~\mathrm{\AA}$   
is expected to facilitate the preferential embedding of Prodan in regions $\mathrm{V}$ and $\mathrm{V}^{\prime}$,
rather than in region $\mathrm{H}$, when Prodan's CoM is located within $z=8-10~\mathrm{\AA}$.
In the presence of ethanol, a decrease in the population of large voids near the membrane center is observed, 
while the overall shape of $P_{\mathrm{V}}\left(z,\sigma\right)$ remains largely changed. 
This reduction may account for the decreased difference
in stability between regions $\mathrm{V+V^{\prime}}$ and $\mathrm{H}$,
as reflected by the decrease in $\Delta G^{\mathrm{V+V^{\prime}},\mathrm{H}}_{\mathrm{MB}}$ (\Fig{fig3}(b)).
\subsection{Electrostatic interaction of Prodan with surrounding environments}
\begin{figure*}[t]
\centering
\includegraphics[width=0.8\linewidth]{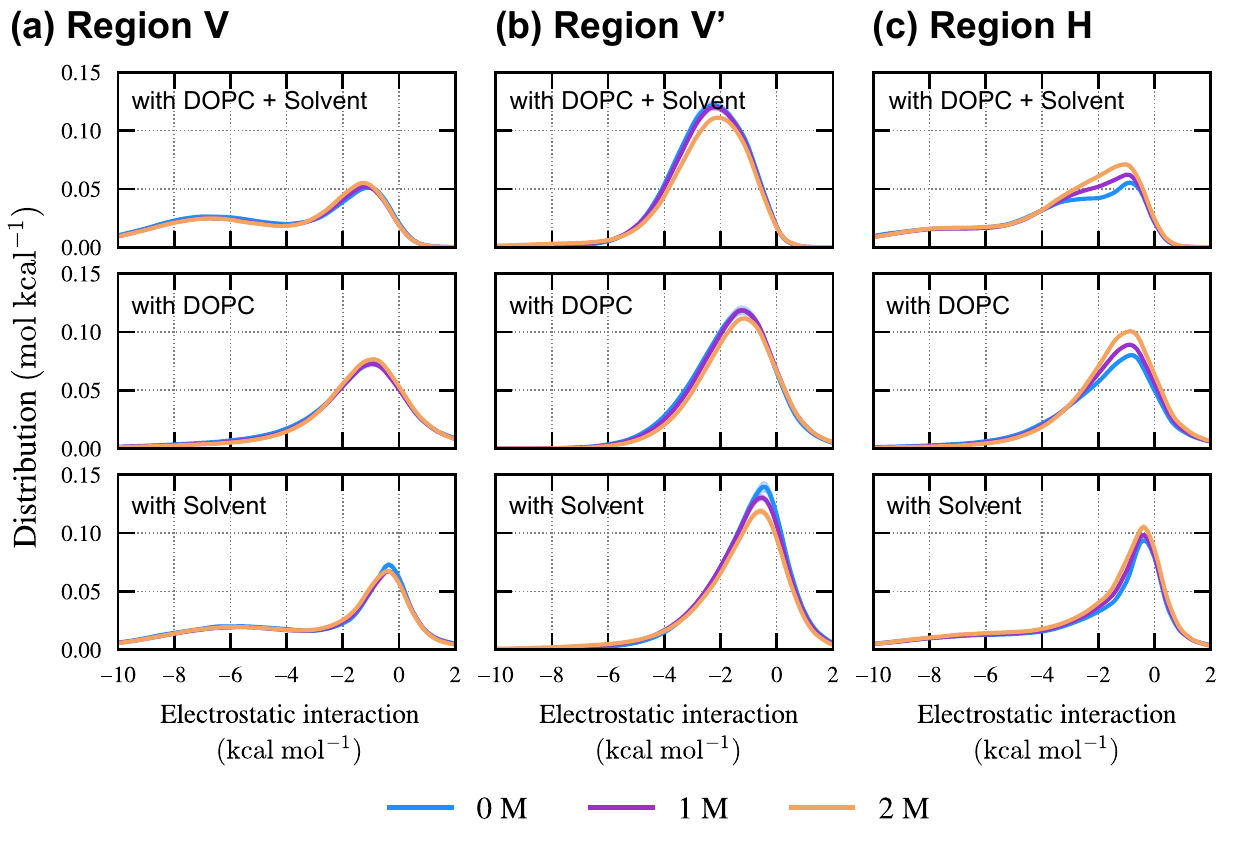}
\caption{Distributions of the electrostatic interaction energies between Prodan 
with its surrounding environments for (a) region $\mathrm{V}$, (b) region $\mathrm{V}^{\prime}$, and (c) region $\mathrm{H}$.
Top, middle, and bottom panels show the distributions of the interactions with DOPC and solvent (water and ethanol), 
with DOPC only, and with solvent only, respectively. 
\label{fig:fig5}}
\end{figure*}
The electrostatic interactions between Prodan and its surrounding environments are elucidated in this subsection.
Given that the electrostatic interactions dominantly contribute to the excitation properties,
analyzing the interaction patterns is beneficial for discussing the connection between the embedded state and the excitation behavior.

In \Fig{fig5}, the distributions of the electrostatic interaction
for different embedded states, $\mathrm{V}$, $\mathrm{V}^{\prime}$, 
and $\mathrm{H}$, are illustrated.
The distributions are calculated from snapshots in which the CoM of Prodan 
is located closer to the membrane center than the average position of the phosphorus atoms in the DOPC molecules.
For region $\mathrm{V}$ (\Fig{fig5}(a)), there are two peaks at $\sim${}$-7~\mathrm{kcal~mol^{-1}}$ and $\sim${}$-1~\mathrm{kcal~mol^{-1}}$, 
and the shape of the distribution is hardly changed upon the addition of ethanol. 
As revealed by the decomposition into contributions from DOPC and solvent (water and ethanol), 
the peak at $\sim$ $-7~\mathrm{kcal~mol^{-1}}$ originates from the solvent.  
Since the propionyl moiety is oriented toward the membrane–solvent interface in region $\mathrm{V}$  
and can strongly interact with the solvent,  
it is reasonable that the solvent contributes to this peak.  
It can be seen that the peak at $\sim${}$-1~\mathrm{kcal~mol^{-1}}$ stems from
both the DOPC and solvent.
Regarding region $\mathrm{V}^{\prime}$ (\Fig{fig5}(b)), 
only a broad distribution centered at $\sim${}$-2~\mathrm{kcal~mol^{-1}}$ is observed.
Since the dimethylamine group is located near the membrane-solvent surface in region $\mathrm{V}^{\prime}$, 
the absence of the peak observed at the more negative value in region $\mathrm{V}$ 
reflects the lower hydrophilicity of the dimethylamine group 
compared with the propionyl group. 
As the ethanol concentration increases,  
the peak of the interaction with DOPC and solvent 
shifts in a positive direction.
This shift is caused by the slight weakening of the interaction with DOPC.
As shown in \Fig{fig5}(c), region $\mathrm{H}$ exhibits a distribution in which the interaction with DOPC and solvent 
is more negative than $-6~\mathrm{kcal~mol^{-1}}$, similar to region $\mathrm{V}$.
Adding ethanol increases the population around $\sim${}$–1~\mathrm{kcal~mol^{-1}}$, and this change clearly originates from DOPC. 
\section{Conclusion}
In this study, 
we investigated the embedding of Prodan within a lipid bilayer composed of 1,2-dioleoyl-\textit{sn}-glycero-3-phosphocholine (DOPC) 
using the molecular dynamics (MD) simulation with the replica-exchange umbrella sampling (REUS) method.
Employing the REUS method enables the evaluation of free-energy landscapes (FELs) for the embedded state with  high statistical reliability.
The response of the embedded states to the addition of ethanol as a cosolvent in the aqueous phase was also examined, given that ethanol is known to distort membrane structures.

The two-dimensional free-energy landscapes (2D-FELs) along the $z$-component of the center of mass (CoM) of Prodan and its orientation identified two stable regions within the membrane, in which the propionyl and dimethylamine groups of Prodan are respectively directed toward the aqueous phase.
These regions were referred to as $\mathrm{V}$ and $\mathrm{V}^{\prime}$, respectively, 
while the less stable region in which the molecular axis of Prodan is oriented perpendicular to the $z$-direction (normal to the membrane surface) 
was denoted as $\mathrm{H}$.
The stabilities of regions $\mathrm{V}$ and $\mathrm{V}^{\prime}$ were similar, 
with $\mathrm{V}^{\prime}$ being slightly more stable.
The one-dimensional FELs (1D-FELs), restricted to a specific region, revealed that the orientational preferences in regions $\mathrm{V}$ and $\mathrm{V}^{\prime}$ are mitigated by the addition of ethanol to the aqueous phase.
The exact decomposition of the FELs into the interaction energy of Prodan with its surrounding environment 
and the many-body entropic contribution showed that the comparable stability of regions 
$\mathrm{V}$ and $\mathrm{V}^{\prime}$ arises from a balance between stabilizing effects 
from the interaction energy contribution and destabilizing effects from the entropic contribution.
The instability of region $\mathrm{H}$,
compared with the composite region of regions $\mathrm{V}$ and $\mathrm{V}^{\prime}$, $\mathrm{V+V^{\prime}}$, 
originated from the interaction energy near the membrane center
and from the entropic contribution around $z \sim${}$8~\mathrm{\AA}$.
The stability difference between region $\mathrm{H}$ and $\mathrm{V+V^{\prime}}$ 
decreased upon the addition of ethanol, due to reduced differences in both the interaction-energy and the entropic contributions.
Furthermore, the void-size distributions within the membrane suggest that
the reduced destabilizing effects from the entropic contribution are correlated with a decrease in large voids near the membrane center.
Such large voids can accommodate bulky molecules or moieties, so their reduction may lead to the relative destabilization of region $\mathrm{V+V^{\prime}}$ when Prodan is located within $z=8-10~\mathrm{\AA}$, 
thereby reducing the stability difference.
The electrostatic interaction of Prodan with its surrounding environment
was stronger in the order $\mathrm{V}^{\prime} > \mathrm{H} > \mathrm{V}$,
a trend primarily determined by its interaction with the solvent.

To further investigate a probe within membranes,
incorporating a theoretical treatment of its electronic structure is an important step.
Practically, the quantum mechanics/molecular mechanics (QM/MM) method
is a useful approach for studying excitation properties 
in heterogeneous environments such as membranes.
Osella \emph{et al.}\cite{Osella_2016} and Wasif \emph{et al.}\cite{wasif2018orientation}
applied this method to investigate the excitation of probes within membranes,
successfully characterizing their spectra.
Recent advances in computing power and molecular dynamics (MD) software\cite{yagi2025high}
have made it possible to combine QM/MM simulations with the REUS method for large-scale systems.
Thus, employing such a technique could 
allow for a more realistic description of the embedded states.
Investigating the effects of different protonation states 
of Prodan on its embedded states is another imporant step, 
especially when the target membrane is associated with the gastrointestinal environment, where the pH ranges from 1.5 to 8. 
The $\mathrm{p}K_{\mathrm{a}}$ of the dimethylamine group in Prodan is $\sim${}$4.5$, 
and hence its predominant protonation state
varies across different regions of the gastrointestinal membrane.
Constant-pH MD methods,\cite{lee2004constant,stern2007molecular,wallace2011continuous,Chen_2015,radak2017constant,de2022constant} 
which enable us to treat multiple protonation states, would be useful 
for analyzing the embedded states of probes within membranes that mimic the gastrointestinal environment. 
Further elucidation using the aforementioned sophisticated techniques
would deepen our understanding of membrane properties 
through a more detailed interpretation of the spectra.
\section*{Supplementary Material}
The supplementary material contains the initial and optimized parameters for the restraint potentials used in the REUS simulations, the free-energy landscapes of Prodan along the $z$-component of the center of mass (CoM) for Prodan, $z$, and the cosine of the tilt angle, $\cos\theta$, and the distributions along the $z$-direction of atoms in phosphate group, glycerol group, and acyl chains. 
\begin{acknowledgments}
This work is supported by the Grants-in-Aid for Scientific Research 
(Nos. 
JP21H05249,     
JP23K27313,     
JP23K26617,     
JP25KJ1759,     
and JP25K17896) 
from the Japan Society for the Promotion of Science, 
the Fugaku Supercomputer Project 
(Nos. JPMXP1020230325 and JPMXP1020230327) and the Data-Driven Material Research Project (No. JPMXP1122714694) from the Ministry of Education, Culture, Sports, Science, and Technology, the Core Research for Evolutional Science and Technology (CREST) 
from Japan Science and Technology Agency (JST) (No. JPMJCR22E3),  
and by Maruho Collaborative Project for Theoretical Pharmaceutics. 
The simulations were conducted using Genkai A at Kyushu University, and Fugaku at RIKEN Center for Computational Science through the HPCI System Research Project
(Project IDs: 
hp250115,      
hp250211,      
hp250227,      
and hp250229). 
\end{acknowledgments}
\section*{Conflict of interest}
The authors have no conflicts to disclose.
\section*{Data Availability}
The data that support the findings of this study are available from the corresponding authors upon reasonable request.
\appendix
\section{Definition of void-size distribution $P_{\mathrm{V}}\left(z,\sigma\right)$\label{sec:def_vsd}}
To analyze the distribution of voids within the membrane, defined as regions not occupied by atoms in the system,  
we employ a numerical technique for estimating void size, 
developed in the field of mesoporous material science.\cite{sarkisov2020materials}
The voids referred to here are called ``pores'' in that field.
However, since ``voids'' is the term used in a recent membrane simulation study,\cite{chipot2016subdiffusion} 
we adopt it in this work.

Let $d_{\mathrm{V},\lambda}\left(z\right)$ denote the size of the $\lambda$th void whose center, ${\bf x}_{\lambda}$, lies 
between $z$ and $z+\Delta z$.
We suppose that all the voids located within this interval satisfy the following conditions.  
\begin{align}
&\forall \lambda \leq N_{\mathrm{V}}\left(z\right)\quad d_{\mathrm{V},\lambda}\left(z\right) =2\min_{i}\left(\left|{\bf x}_{\lambda}-{\bf r}_{i}\right|-\dfrac{\sigma_{i}}{2}\right),
\label{dV_def}
\\
&\forall \lambda,\mu \leq N_{\mathrm{V}}\left(z\right), 
\lambda \neq \mu \quad  \left| \mathbf{x}_\lambda - \mathbf{x}_\mu \right| > \dfrac{1}{2}d_{\mathrm{V},\mu}(z).
\label{void_separation}
\\
&\forall\lambda\leq N_{\mathrm{V}}\left(z\right),\forall i\leq N_{\mathrm{atom}} \quad\left|{\bf x}_{\lambda}-{\bf r}_{i}\right|>\dfrac{\sigma_{i}}{2}.
\label{void_outside_atoms}
\end{align}
Here, ${\bf r}_{i}$ is the position of $i$th atom in the system, and $\sigma_{i}$ is its diameter.
$N_{\mathrm{V}}\left(z\right)$ is the number of voids whose centers are located between $z$ and $z+\Delta z$, 
and $N_{\mathrm{atom}}$ is the number of atoms in the system.
$d_{\mathrm{V},\lambda}\left(z\right)$ represents the maximum diameter of the sphere centered at ${\bf x}_{\lambda}$ that does not contain any atoms.
\Eqs{void_separation}{void_outside_atoms} ensure 
that no void contains the center of any other void, 
and that all void centers are located outside the atoms in the system, respectively.
Then, let us define the unnormalized void density as 
\begin{align}
\rho_{\mathrm{V}}\left(z,\sigma\right) & =\Braket{\dfrac{1}{\Delta z}\sum_{i=1}^{N_{\mathrm{V}}\left(z\right)}\delta\left(\sigma-d_{\mathrm{V},i}\left(z\right)\right)}.
\end{align}
Normalization is introduced as
\begin{align}
P_{\mathrm{V}}\left(z,\sigma\right) & =\rho_{\mathrm{V}}\left(z,\sigma\right)\left({\displaystyle \int_{0}^{\mathrm{z_{\mathrm{P}}}}dz\int_{0}^{\infty}d\sigma}\,\rho_{\mathrm{V}}\left(z,\sigma\right)\right)^{-1}.
\end{align}
Here, $z_{\mathrm{P}}$ is the average $z$-coordinate of the phosphorus atoms in DOPC molecules, measured from the membrane center ($z=0$).

The code for computing $P_{\mathrm{V}}\left(z\right)$ is implemented in PoreBlazer v4.0.\cite{sarkisov2020materials}
We use $\sigma$ parameters of the Lennard-Jones interaction in the CHARMM36 force field\cite{klauda2010update}. 
\bibliography{prodan}
\clearpage
\widetext

\def\thesection{S\arabic{section}}
\setcounter{section}{0}
\renewcommand{\theequation}{S\arabic{equation}}
\setcounter{equation}{0}
\renewcommand{\thefigure}{S\arabic{figure}}
\setcounter{figure}{0}
\renewcommand{\thetable}{S\arabic{table}}
\setcounter{table}{0}
\renewcommand{\thepage}{S\arabic{page}}
\setcounter{page}{0}

\begin{center}
\Large \bf
Supplement for ``Characterizing the embedded states of a fluorescent probe within a lipid bilayer using molecular dynamics simulations''
\end{center}
\begin{table}[h]
\vspace*{-\intextsep}
\centering
\caption{Initial parameters for the restraint potentials for the REUS simulation ($V_{i}\left(z\right) = k_{i}\left(z-z_{i}\right)^{2}$) in the case of the pure water phase. Subscript $i$ represent the replica index. \label{tab:REUS_initial}}
\renewcommand{\arraystretch}{1.2}
\begin{tabular}{ccccccccc}
 \hline \hline
 Replica index   & 1 & 2 & 3& 4 &5 & 6 & 7&8\\
 \hline 
 $k_i$ (kcal mol$^{-1}$ $\text{\AA}{}^{-2}$)&1.0 &1.0 &1.0 &1.0&1.0 &1.0& 1.0 &1.0\\ 
 $z_{i}\;(\text{\AA})$ & 0.00 & 1.00  & 2.00 & 3.00 & 4.00 & 5.00 & 6.00 & 7.00 \\
 \hline
 \hline
 Replica index & 9 & 10 & 11& 12 &13&14&15&16\\
 \hline 
 $k_i$ (kcal mol$^{-1}$ $\text{\AA}{}^{-2}$)&1.0 &1.0 &1.0 &1.0&1.0 &1.0& 1.0 &1.0\\ 
 $z_{i}\;(\text{\AA})$ & 8.00 & 9.00 & 10.00 & 11.00 & 12.00 & 13.00 & 14.00 & 15.00 \\
 \hline
 \hline
 Replica index & 17 & 18 & 19& 20 & 21& 22& 23& 24\\
 \hline 
 $k_i$ (kcal mol$^{-1}$ $\text{\AA}{}^{-2}$)&1.0 &1.0 &1.0 &1.0&1.0 &1.0& 1.0 &1.0\\ 
 $z_{i}\;(\text{\AA})$ & 16.00 & 17.00 & 18.00 & 19.00 & 20.00 & 21.00 & 22.00 & 23.00 \\
 \hline
 \hline
 Replica index & 25 & 26 & 27& 28 & 29& 30& 31& 32\\
 \hline 
 $k_i$ (kcal mol$^{-1}$ $\text{\AA}{}^{-2}$)&1.0 &1.0 &2.5 &2.5&2.5 &2.5& 2.5 &2.5\\ 
 $z_{i}\;(\text{\AA})$ & 24.00 & 25.00 & 25.60 & 26.20 & 26.80 & 27.40 & 28.00 & 28.60 \\
 \hline
\end{tabular}
\end{table}
\begin{table}[h]
\centering
\vspace*{-\intextsep}
\caption{Optimized parameters for the restraint potentials used in the REUS simulation ($V_{i}\left(z\right) = k_{i}\left(z-z_{i}\right)^{2}$) for the pure water phase. The initial parameters are listed in \Table{REUS_initial}. 
Subscript $i$ represent the replica index. \label{tab:REUS_0M_param}}
\renewcommand{\arraystretch}{1.2}
\begin{tabular}{ccccccccc}
 \hline \hline
 Replica index   & 1 & 2 & 3& 4 &5 & 6 & 7&8\\
 \hline 
 $k_i$ (kcal mol$^{-1}$ $\text{\AA}{}^{-2}$)&1.0 &1.0 &1.0 &1.0&1.0 &1.0& 1.0 &1.0\\ 
 $z_{i}\;(\text{\AA})$ & 0.00 & 0.89  & 1.78 & 2.68 & 3.47 & 4.35 & 5.22 & 6.20\\
 \hline
 \hline
 Replica index & 9 & 10 & 11& 12 &13&14&15&16\\
 \hline 
 $k_i$ (kcal mol$^{-1}$ $\text{\AA}{}^{-2}$)&1.0 &1.0 &1.0 &1.0&1.0 &1.0& 1.0 &1.0\\ 
 $z_{i}\;(\text{\AA})$ & 7.22 & 8.09 & 8.97 & 9.87 & 10.79 & 11.68 & 12.59 & 13.51\\
 \hline
 \hline
 Replica index & 17 & 18 & 19& 20 & 21& 22& 23& 24\\
 \hline 
 $k_i$ (kcal mol$^{-1}$ $\text{\AA}{}^{-2}$)&1.0 &1.0 &1.0 &1.0&1.0 &1.0& 1.0 &1.0\\ 
 $z_{i}\;(\text{\AA})$ & 14.43 & 15.36 & 16.25 & 17.09 & 18.12 & 18.98 & 19.82 & 20.76\\
 \hline
 \hline
 Replica index & 25 & 26 & 27& 28 & 29& 30& 31& 32\\
 \hline 
 $k_i$ (kcal mol$^{-1}$ $\text{\AA}{}^{-2}$)&1.0 &1.0 &2.5 &2.5&2.5 &2.5& 2.5 &2.5\\ 
 $z_{i}\;(\text{\AA})$ & 21.65 & 22.55 & 23.14 & 23.74 & 24.30 & 24.85 & 25.50 & 26.09\\
 \hline
\end{tabular}
\end{table}
\begin{table}[h]
\centering
\vspace*{-\intextsep}
\caption{Optimized parameters for the restraint potentials used in the REUS simulation ($V_{i}\left(z\right) = k_{i}\left(z-z_{i}\right)^{2}$) for the 1 M ethanol aqueous phase. The initial parameters are taken from the optimized ones for the pure water case (\Table{REUS_0M_param}). 
Subscript $i$ represent the replica index.}
\renewcommand{\arraystretch}{1.2}
\begin{tabular}{ccccccccc}
 \hline \hline
 Replica index   & 1 & 2 & 3& 4 &5 & 6 & 7&8\\
 \hline 
 $k_p$ (kcal mol$^{-1}$ $\text{\AA}{}^{-2}$)&1.0 &1.0 &1.0 &1.0&1.0 &1.0& 1.0 &1.0\\ 
 $z_{i}\;(\text{\AA})$ & 0.00 & 0.87  & 1.76 & 2.63 & 3.52 & 4.50 & 5.39 & 6.32\\
 \hline
 \hline
 Replica index & 9 & 10 & 11& 12 &13&14&15&16\\
 \hline
 $k_i$ (kcal mol$^{-1}$ $\text{\AA}{}^{-2}$)&1.0 &1.0 &1.0 &1.0&1.0 &1.0& 1.0 &1.0\\ 
 $z_{i}\;(\text{\AA})$ &7.17 &8.13 & 9.07 & 9.98 & 10.93 & 11.79 & 12.53 & 13.38\\
 \hline
 \hline
 Replica index & 17 & 18 & 19& 20 & 21& 22& 23& 24\\
 \hline 
 $k_i$ (kcal mol$^{-1}$ $\text{\AA}{}^{-2}$)&1.0 &1.0 &1.0 &1.0&1.0 &1.0& 1.0 &1.0\\ 
 $z_{i}\;(\text{\AA})$ & 14.22 & 15.13 & 16.05 & 16.93 & 17.69 & 18.63 & 19.63 & 20.62\\
 \hline
 \hline
 Replica index & 25 & 26 & 27& 28 & 29& 30& 31& 32\\
 \hline 
 $k_i$ (kcal mol$^{-1}$ $\text{\AA}{}^{-2}$)&1.0 &1.0 &2.5 &2.5&2.5 &2.5& 2.5 &2.5\\ 
 $z_{i}\;(\text{\AA})$ & 21.38 & 22.27 & 22.79 & 23.34 & 23.90 & 24.51 & 25.11 & 25.66\\
 \hline
\end{tabular}
\end{table}
\begin{table}[h]
\centering
\vspace*{-\intextsep}
\caption{Optimized parameters for the restraint potentials used in the REUS simulation ($V_{i}\left(z\right) = k_{i}\left(z-z_{i}\right)^{2}$) for the 2 M ethanol aqueous phase. The initial parameters are taken from the optimized ones for the pure water case (\Table{REUS_0M_param}). 
Subscript $i$ represent the replica index.}
\renewcommand{\arraystretch}{1.2}
\begin{tabular}{ccccccccc}
 \hline \hline
 Replica index   & 1 & 2 & 3& 4 &5 & 6 & 7&8\\
 \hline 
 $k_p$ (kcal mol$^{-1}$ $\text{\AA}{}^{-2}$)&1.0 &1.0 &1.0 &1.0&1.0 &1.0& 1.0 &1.0\\ 
 $z_{i}\;(\text{\AA})$ & 0.00 & 0.85  & 1.61 & 2.57 & 3.44 & 4.37 & 5.21 & 6.16\\
 \hline
 \hline
 Replica index & 9 & 10 & 11& 12 &13&14&15&16\\
 \hline 
 $k_i$ (kcal mol$^{-1}$ $\text{\AA}{}^{-2}$)&1.0 &1.0 &1.0 &1.0&1.0 &1.0& 1.0 &1.0\\ 
 $z_{i}\;(\text{\AA})$ & 7.04 & 8.00 & 8.86 & 9.68 & 10.50 & 11.47 & 12.48 & 13.31\\
 \hline
 \hline
 Replica index & 17 & 18 & 19& 20 & 21& 22& 23& 24\\
 \hline 
 $k_i$ (kcal mol$^{-1}$ $\text{\AA}{}^{-2}$)&1.0 &1.0 &1.0 &1.0&1.0 &1.0& 1.0 &1.0\\ 
 $z_{i}\;(\text{\AA})$ & 14.25 & 15.07 & 16.13 & 17.05 & 17.94 & 18.88 & 19.87 & 20.80\\
 \hline
 \hline
 Replica index & 25 & 26 & 27& 28 & 29& 30& 31& 32\\
 \hline 
 $k_i$ (kcal mol$^{-1}$ $\text{\AA}{}^{-2}$)&1.0 &1.0 &2.5 &2.5&2.5 &2.5& 2.5 &2.5\\ 
 $z_{i}\;(\text{\AA})$ & 21.69 & 22.58 & 23.16 & 23.82 & 24.35 & 25.00 & 25.67 & 26.23\\
 \hline
\end{tabular}
\end{table}
\clearpage
\begin{figure}[t]
\centering
\vspace*{-\intextsep}  
\includegraphics[width=1.0\linewidth]{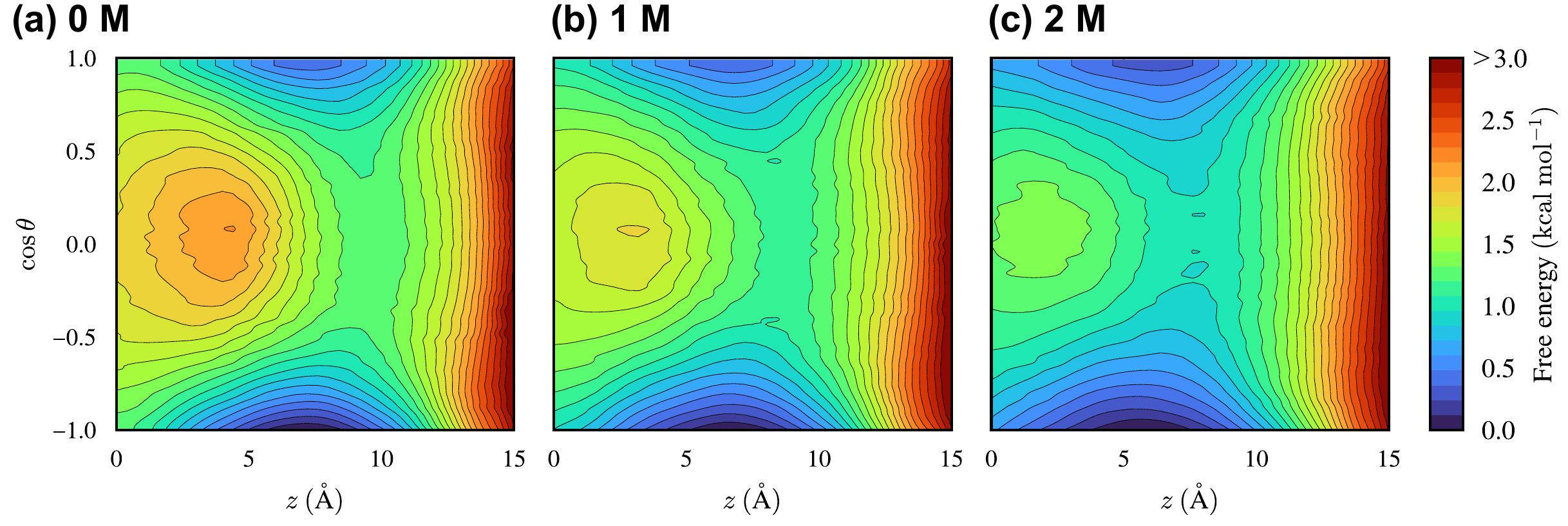}
\caption{Free-energy landscapes on the $z$-component of the center of mass (CoM) for Prodan, $z$, and the cosine of the title angle, $\cos\theta$, for (a) pure water, (b) 1 M ethanol, and (c) 2 M ethanol aqueous phases.}
\end{figure}
\begin{figure}[t]
\centering
\vspace*{-\intextsep}  
\includegraphics[width=1.0\linewidth]{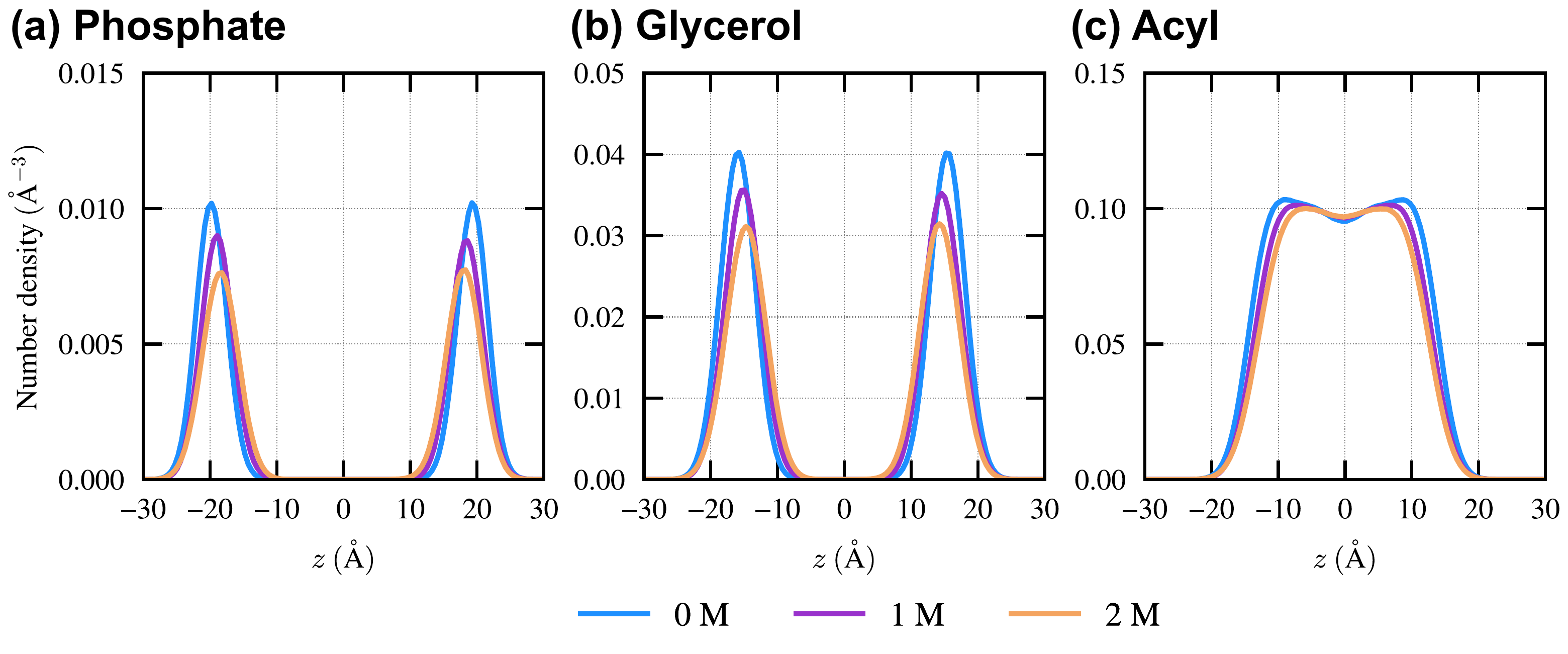}
\caption{Distributions along the $z$-direction of atoms in (a) phosphate group, (b) glycerol group, and (c) acyl chains.
}
\end{figure}
\end{document}